\documentclass[11pt]{article}
\usepackage[utf8]{inputenc}
\usepackage{siunitx}
\usepackage[letterpaper, margin=1in]{geometry}
\usepackage{fancyhdr}
\usepackage{mdframed}
\usepackage{amssymb}
\usepackage{amsmath}
\usepackage{graphicx}
\usepackage{float}
\usepackage{parskip}
\usepackage{hyperref}
\usepackage[bottom]{footmisc}
\usepackage{authblk}

\usepackage[sorting=none,bibstyle=numeric,hyperref=true]{biblatex}
\usepackage{tikz}
\usepackage{ifthen}
\usetikzlibrary{arrows}
\usepackage{tikz-3dplot}

\author[1]{A.~W.~Nelsen}
\author[1]{E.~G.~Ballantyne}
\author[1]{R.~E.~Calvert}
\author[1]{C.~B.~Crawford}
\author[2]{G.~L.~Greene}
\author[1]{S.~E.~Vickers}
\author[3]{F.~E.~Wietfeldt}

\affil[1]{University of Kentucky, Lexington, KY, USA}
\affil[2]{University of Tennessee, Knoxville, TN, USA}
\affil[3]{Tulane University, New Orleans, LA, USA}

\title{Geometric Optimization of a Neutron Detector for In-Flight Measurement of the Neutron Lifetime}
\date{}

\begin{document}

\maketitle

\section{Introduction}
The neutron lifetime is essential input in determining initial abundances of light elements in Big Bang Nucleosynthesis. The lifetime, together with decay correlations, can also be used to extract the vector and axial vector weak coupling constants ($G_V$ and $G_A$) of the neutron. Under the Conserved Vector Current hypothesis, $G_V$ of the neutron gives access to the CKM matrix element $V_{ud}$ without systematic uncertainty due to nuclear structure~\cite{Beck:2019xye,Broussard:2018cbi}.

There are two standard techniques to measure the neutron lifetime. Originally, the beam method was used, which measures the fractional decay rate of in-flight neutrons on a cold neutron beamline.
With the advent of ultracold neutrons, which could be stored for time scales comparable to the neutron lifetime, experiments started using the bottle method, where the fraction of neutrons remaining in a storage volume is measured as a function of time.

Measurement of the lifetime of the free neutron using the beam method~\cite{spi88,byr96,yue13,nag18} has an 8.7~s (4$\sigma$) discrepancy with UCN bottle measurements~\cite{mam93,ser05,pic10,ste12,arz15,ser18,pat18,ezh18}.
Two possible explanations for the discrepancy are missing decay protons, resulting in a perceived longer lifetime in the beam method, or additional neutron losses, such as wall losses or decays to exotic particles in the bottle method~\cite{Cline:2018ami,Fornal:2018eol}.
The goal of the BL3 experiment is to improve the statistical error of this measurement and help rule out the former systematic uncertainties as an explanation for the discrepancy. In this paper we report optimization on the detector geometry to minimize the uncertainty in the neutron fluence due to variations in neutron beam profile coupled with nonuniform detection efficiency.

\subsection{Experimental Setup}
\begin{figure}[t!]
    \centering
    \includegraphics[width=0.8\linewidth]{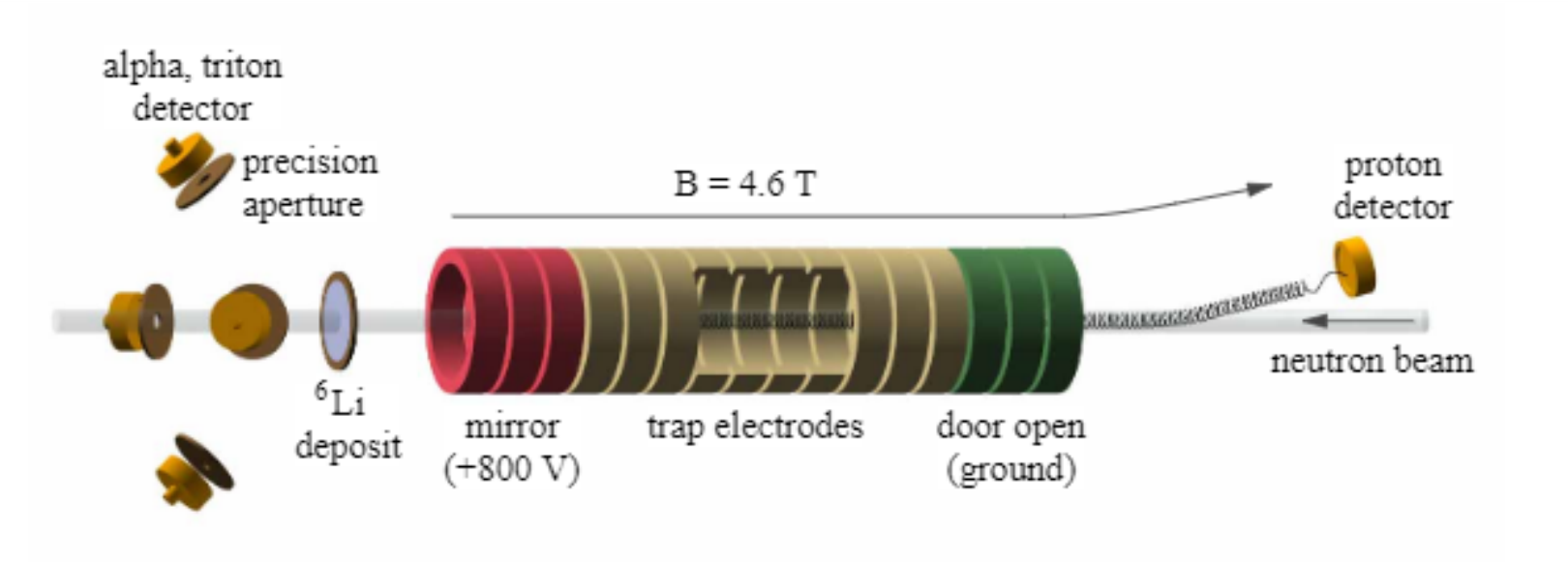}
    \caption{The beam trap and flux detector configuration. A beam of cold neutrons passes through the trap (in through the ``door'' and out through the mirror). Decay protons are released from the trap at regular intervals and directed to the detector. The beam leaves the trap and passes through the $^6$Li foil. Alpha and triton products are detected in the surrounding detectors.}
    \label{fig:beam_trap}
\end{figure}

In the beam method experiment, shown in Fig.~\ref{fig:beam_trap}, a beam of cold neutrons is passed through a known volume, decaying at a rate of $R_p$. All decay protons are held in a Penning trap and detected with an efficiency $\varepsilon_p\sim1$.
The neutron beam passes through a thin foil (typically $^6$LiF on a substrate) downstream of the detector volume. Some neutrons are captured by the $^6$Li nuclei and the alpha and triton products are captured with an efficiency $\varepsilon_{th}$ for neutrons at the reference thermal velocity $v_{th} = 2200$ m/s~\cite{wiet14} (see also \cite[p.~1178]{wiet11}).

Because the neutron energy is well below any resonance in the ${^6}$Li, the capture cross section has a pure $1/v$ dependence. Since the proton decay rate $R_p$ also directly depends on the neutron flux weighted by the time $\Delta t=L_t/v$ each neutron spends in the trap, the $1/v$ dependence in the neutron detection rate $R_n$ cancels out the same in $R_p$.  This dramatically reduces the systematic error due to uncertainty of the neutron spectrum. In terms of these quantities, the lifetime of the neutron is~\cite{wiet14}
\begin{equation}    \label{eq:lifetime}
    \tau_n = \frac{R_n \varepsilon_p L_t}{R_p \varepsilon_{th} v_{th}}.
\end{equation}
The effective trap length $L_t$ is particularly challenging since the effective length, including end effects, is not well known, so a series of measurements are conducted with different trap lengths (see Fig.~\ref{fig:beam_trap}). The difference in these measurements is used to cancel edge effects of the trap length.

Since the neutron counting rate enters linearly into the neutron lifetime, it is essential to have a well-characterized neutron flux detector with flat response function $\Omega(\delta,\epsilon)$, where $\delta$ and $\epsilon$ are the position of neutron capture in the $^6$Li foil plane (Fig.~\ref{fig:2D_geometry}, \ref{fig:3D_geometry}). This minimizes dependence of the neutron detection efficiency on the profile of the neutron beam.

In this paper, we eliminate leading order terms in the Taylor expansion of $\Omega(\delta,\epsilon;\{P\})$ about $\delta,\epsilon=0$, by tuning a set $\{P\}$ of geometric parameters. These parameters include the polar angle $\Theta$ of the center of the detector with respect to the normal of the $^6$Li foil plane, and the tilt angle $\Gamma$ of the detector normal away from perpendicular to the origin (see Fig.~\ref{fig:2D_geometry}).
Furthermore, a non-uniform $^6$LiF deposit is considered, introducing another parameter $\zeta$, the height that the point-like evaporator is placed above the substrate surface during the coating process, as a fraction of the distance $r$ from the center of the foil to the detector.

\section{Two Detectors} \label{sec:2det}
    
    \begin{figure}[b!]
        \centering
        \tdplotsetmaincoords{90}{0}
        \begin{tikzpicture}[scale=2,tdplot_main_coords]
        
            \coordinate (O) at (0,0,0);
            \tdplotsetcoord{P}{2.5}{45}{0};
            
            \tdplotsetcoord{X}{1.5}{90}{0};
            
            \draw[thick,->] (0,0,0) -- (2.5,0,0) node[anchor=north east]{$x$};
            \draw[thick,->] (0,0,0) -- (0,0,2.5) node[anchor=south]{$z$};
            
            \draw[->,red] (O) -- (P);   
                \tdplotsetcoord{p}{1.25}{45}{0};    
                \node at (p)[anchor=south east]{$\vec{r}$};
                
            \tdplotsetthetaplanecoords{0};
            \tdplotdrawarc[tdplot_rotated_coords]{(0,0,0)}{0.5}{0}{45}{anchor=south}{$\Theta$}; 
                
            \draw[->,blue] (O) -- (X);  
                \tdplotsetcoord{x}{0.75}{90}{0};
                \node at (x) [anchor=south]{$\vec{x}$};
            \filldraw[color=black,fill=blue] (X) circle (1pt);
            
            \draw[->,orange] (X) -- (P);    
                \tdplotsetcoord{d}{1.81}{64.8}{0};
                \node at (d) [anchor=west]{$\vec{d}$};

            \tdplotsetrotatedcoordsorigin{(P)};     
            \tdplotsetrotatedcoords{0}{-25}{0}; 
            
            \draw[tdplot_rotated_coords] (0,0,-0.5) -- (0,0,0.5);  
                \draw[->,dashed,tdplot_rotated_coords] (0,0,0) -- (0.5,0,0) node [anchor=north]{$\hat{A}$};
                
            \tdplotsetrotatedcoords{0}{-45}{0}; 
            \draw[->,tdplot_rotated_coords,red,dashed] (0,0,0) -- (0.5,0,0);
                \tdplotsetcoord{r}{3}{45}{0};
                \node at (r) [anchor=south]{$\hat{r}$};
                
            \tdplotsetrotatedthetaplanecoords{0};
            \tdplotdrawarc[tdplot_rotated_coords]{(0,0,0)}{0.45}{90}{110}{anchor=south west}{$\Gamma$}; 
            
            \tdplotsetrotatedcoords{0}{8.6}{0}; 
            \draw[->,tdplot_rotated_coords,dashed,orange] (0,0,0) -- (0,0,0.5);
                \tdplotsetcoord{d1}{2.88}{36.4}{0};
                \node at (d1) {$\hat{d}$};
            
        \end{tikzpicture}
        \caption{Detector geometry in two dimensions. The detector is located a distance $r$ from the center of the foil plane at an angle $\Theta$. An arbitrary neutron decay point along the $x$ axis is given by the blue dot at $\delta\equiv x/r$. The distance from that point to the center of the detector is given by $d$.}
        \label{fig:2D_geometry}
    \end{figure}
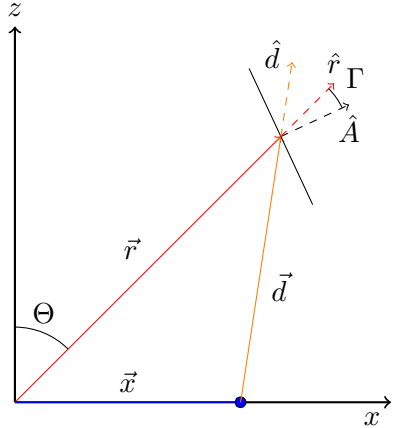
    
\begin{figure}[t!]
        \centering
        \includegraphics[width=0.45\textwidth]{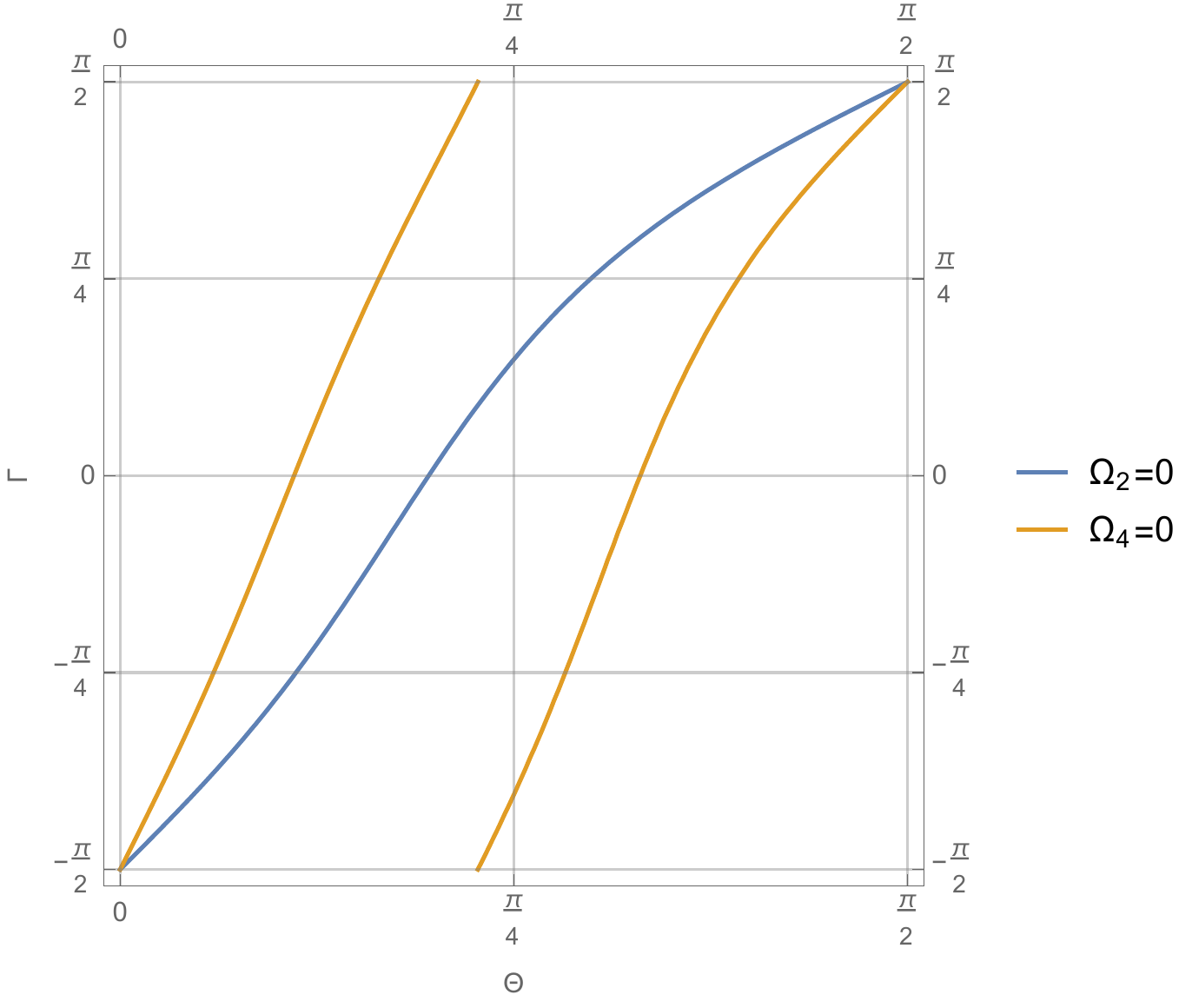}
        \caption{Contour plot of expansion coefficients $\Omega_2=0$ and $\Omega_4=0$ in the parameter space $(\Theta,\Gamma)$ for 2-dimensional detector geometry. The only intersection points are at $\Theta=\pi/2$, $\Gamma=\pi/2$ and $\Theta=0$, $\Gamma=-\pi/2$, which are trivial solutions.}
        \label{fig:2D_contour}
    \end{figure}

    We first consider the two detector case; at distances $\pm x$ from the origin (see Fig.~\ref{fig:2D_geometry}). This configuration reduces to a two dimensional problem, except for the area of the detectors, assumed to be small enough such that we can neglect variations in distance from the source across the area.
    The average solid angle of this configuration is\footnotemark
    \begin{equation}    \label{eq:2D_omega}
        \Omega(\Theta,\Gamma) = \frac{1}{2} \left[ \frac{\cos\Gamma - \delta\sin(\Theta+\Gamma)}{(1 + \delta^2-2\delta\sin\Theta)^{3/2}} + \frac{\cos\Gamma + \delta\sin(\Theta+\Gamma)}{(1 + \delta^2+2\delta\sin\Theta)^{3/2}}\right]
    \end{equation}
    where $\delta\equiv x/r$.\footnotetext[1]{For a brief derivation, see Sec.~\ref{ssec:omega}.}
    Expanding the solid angle to fourth order about $\delta=0$, the odd orders of $\delta$ vanish by symmetry.
    The second order and fourth order coefficients ($\Omega_2$ and $\Omega_4$, respectively), which are functions of $\Theta$ and $\Gamma$, is simultaneously eliminated, yielding a system of equations.
    Fig.~\ref{fig:2D_contour} indicates only two optimal solutions: ($\Theta=0,\Gamma=-\pi/2$) and ($\Theta=\pi/2,\Gamma=\pi/2$). Both solutions are trivial since they cause $\Omega$ to vanish.
    
\begin{figure}[b]
        \centering
        \includegraphics[width=0.55\textwidth]{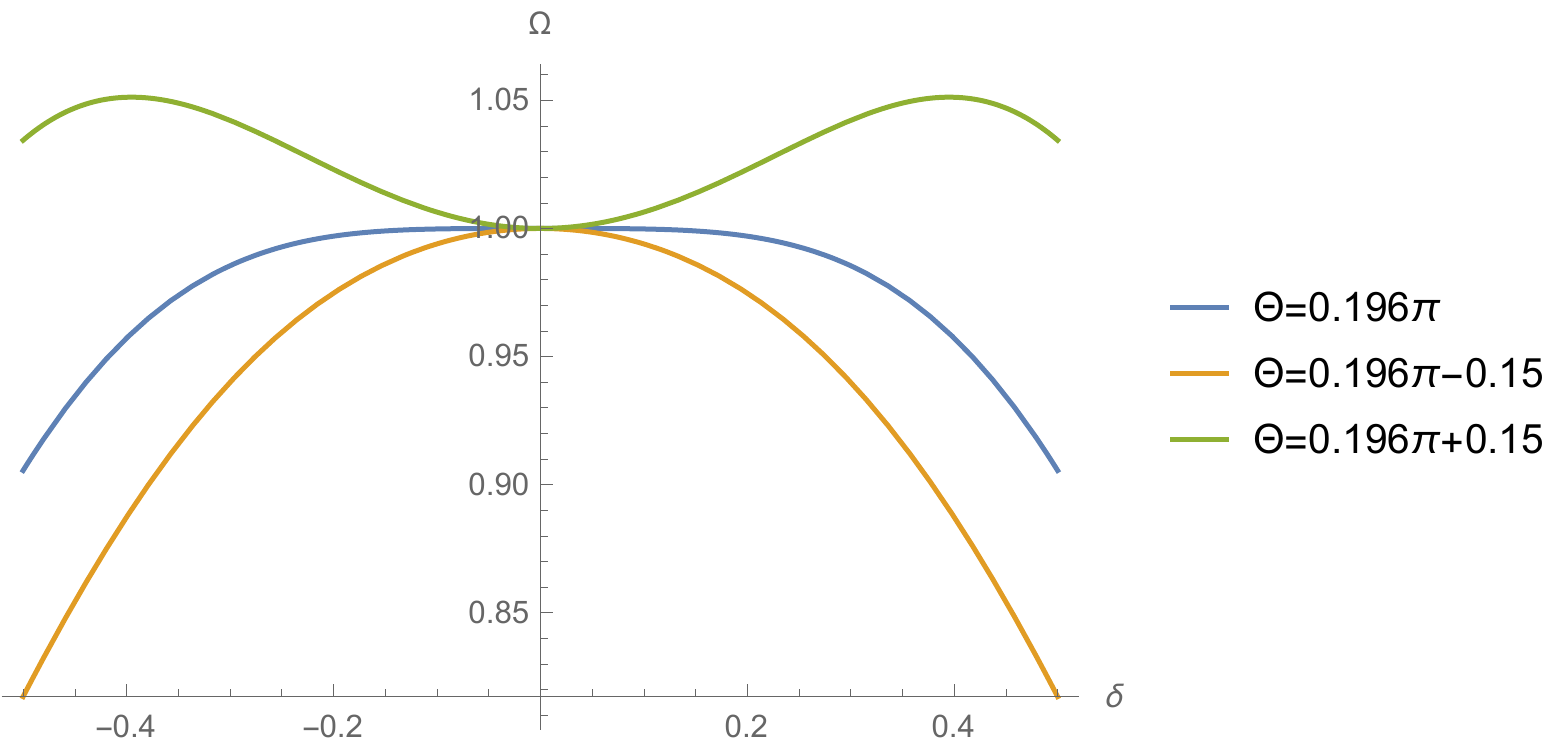}
        \caption{Comparison of $\Theta$'s around the optimal angle ($\Theta= 35.26^{\circ}= 0.196\pi$). For $\delta=0.15$, there is a change in efficiency of $0.10\%$.}
        \label{fig:theta_opt_2D}
    \end{figure}

    To remedy the problem, one parameter must be constrained. $\Gamma=0$ was chosen as this maximizes $\Omega_0=\cos\Gamma$. Solving only $\Omega_2(\Theta,0)=0$ yields $\Theta= 35.26^{\circ}$.
    A comparison with neighboring values of $\Theta$ shows that this optimal $\Theta$ makes $\Omega$ more uniform (Fig.~\ref{fig:theta_opt_2D}). Fig.~\ref{fig:theta_opt_2D} shows non-uniformity of less than 0.72\% in the efficiency until $\delta=0.25$. For a beam of width 3 cm and a point-like detector 20 cm away, $\delta=0.15$, giving a maximal change in efficiency of $0.10\%$.
    
\section{Four Detectors}  \label{sec:4det}

    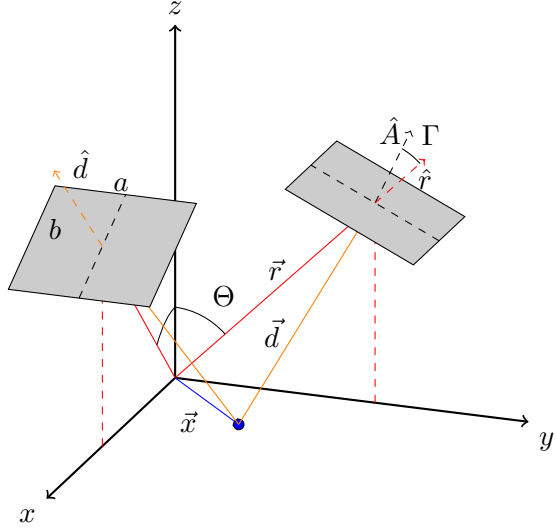
\begin{figure}[t!]
        \centering
        \tdplotsetmaincoords{70}{110}   
        \begin{tikzpicture}[scale=2,tdplot_main_coords]
        
            \coordinate (O) at (0,0,0);
            \tdplotsetcoord{P}{2}{45}{0}
            \tdplotsetcoord{Q}{2}{45}{90}
            
            \tdplotsetcoord{X}{1}{90}{45}   
        
            \draw[thick,->] (0,0,0) -- (2.5,0,0) node[anchor=north east]{$x$};
            \draw[thick,->] (0,0,0) -- (0,2.5,0) node[anchor=north west]{$y$};
            \draw[thick,->] (0,0,0) -- (0,0,2.5) node[anchor=south]{$z$};
            
            \draw[->,red] (O) -- (P);       
                \draw[dashed,red] (Px) -- (P);  
                \tdplotsetthetaplanecoords{0};   
                \tdplotdrawarc[tdplot_rotated_coords]{(0,0,0)}{0.5}{0}{45}{}{}; 
            \draw[->,red] (O) -- (Q);       
                \node at (0,0.71,0.71) [anchor=south]{$\vec{r}$};
                \draw[dashed,red] (Qy) -- (Q);  
                \tdplotsetthetaplanecoords{90}  
                \tdplotdrawarc[tdplot_rotated_coords]{(0,0,0)}{0.5}{0}{45}{anchor=south west}{$\Theta$} 
                
            \draw[->,blue] (O) -- (X);
                \tdplotsetcoord{x}{0.5}{90}{45};  
                \node at (x) [anchor=north east]{$\vec{x}$};
            \filldraw[color=black,fill=blue] (X) circle (1pt);

            \tdplotsetcoord{d}{1}{50}{77};
            
            \draw[->,orange] (X) -- (P);
            \draw[->,orange] (X) -- (Q);
                \node at (d) [anchor=north]{$\vec{d}$};
                
            \tdplotsetrotatedcoordsorigin{(Q)};     
            \tdplotsetrotatedcoords{90}{25}{-90};     
            \filldraw[tdplot_rotated_coords,draw=black,fill=black!20] (0.5,0.5,0) -- (0.5,-0.5,0) -- (-0.5,-0.5,0) -- (-0.5,0.5,0) -- cycle;
            \draw[tdplot_rotated_coords,dashed] (0,0.5,0) -- (0,-0.5,0);
                \draw[->,dashed,tdplot_rotated_coords] (0,0,0) -- (0,0,0.6) node[anchor=east]{$\hat{A}$};
            
            \tdplotsetrotatedcoords{90}{-45}{-90};     
            \tdplotsetrotatedthetaplanecoords{90};   
            \tdplotdrawarc[tdplot_rotated_coords]{(0,0,0)}{0.45}{90}{70}{anchor=south west}{$\Gamma$}; 
                
            \tdplotsetrotatedcoords{90}{-45}{-90};     
            \draw[->,red,dashed,tdplot_rotated_coords] (0,0,0) -- (0,0.5,0);
                \tdplotsetcoord{q}{2.5}{45}{90};
                \node at (q) [anchor=north]{$\hat{r}$};   
                
            \tdplotsetrotatedcoordsorigin{(P)};     
            \tdplotsetrotatedcoords{0}{25}{0};  
            \filldraw[tdplot_rotated_coords,draw=black,fill=black!20] (0.5,0.5,0) -- (-0.5,0.5,0) -- (-0.5,-0.5,0) -- (0.5,-0.5,0) -- cycle;
            \draw[tdplot_rotated_coords,dashed] (0.5,0,0) -- (-0.5,0,0);
                \tdplotsetcoord{a}{2.2213}{38.929}{0};
                \node at (a) [anchor=south west]{$a$};
                \tdplotsetcoord{b}{2.2213}{45}{-6.071};
                \node at (b) [anchor=east]{$b$};
            
            \tdplotsetrotatedcoordsorigin{(P)};
            \tdplotsetrotatedcoords{0}{45}{45};  
            \draw[tdplot_rotated_coords,dashed,orange,->] (0,0,0) -- (-0.42,0,0.47);
                \tdplotsetcoord{dhat}{2.7}{43}{0};
                \node at (dhat) [anchor=south]{$\hat{d}$};

        \end{tikzpicture}
        \caption{Detector geometry in three dimension with an isotropic source in the $(\delta\equiv x/r,\epsilon\equiv y/r)$ plane. The common detector angle $\Theta$ and tilt angle $\Gamma$ are in the same plane for each detector; $xz$ plane for detector 1, $yz$ plane for detector 2, etc. (other two not shown).}
        \label{fig:3D_geometry}
    \end{figure}
    
\begin{figure}[b]
        \centering
        \includegraphics[width=0.35\textwidth]{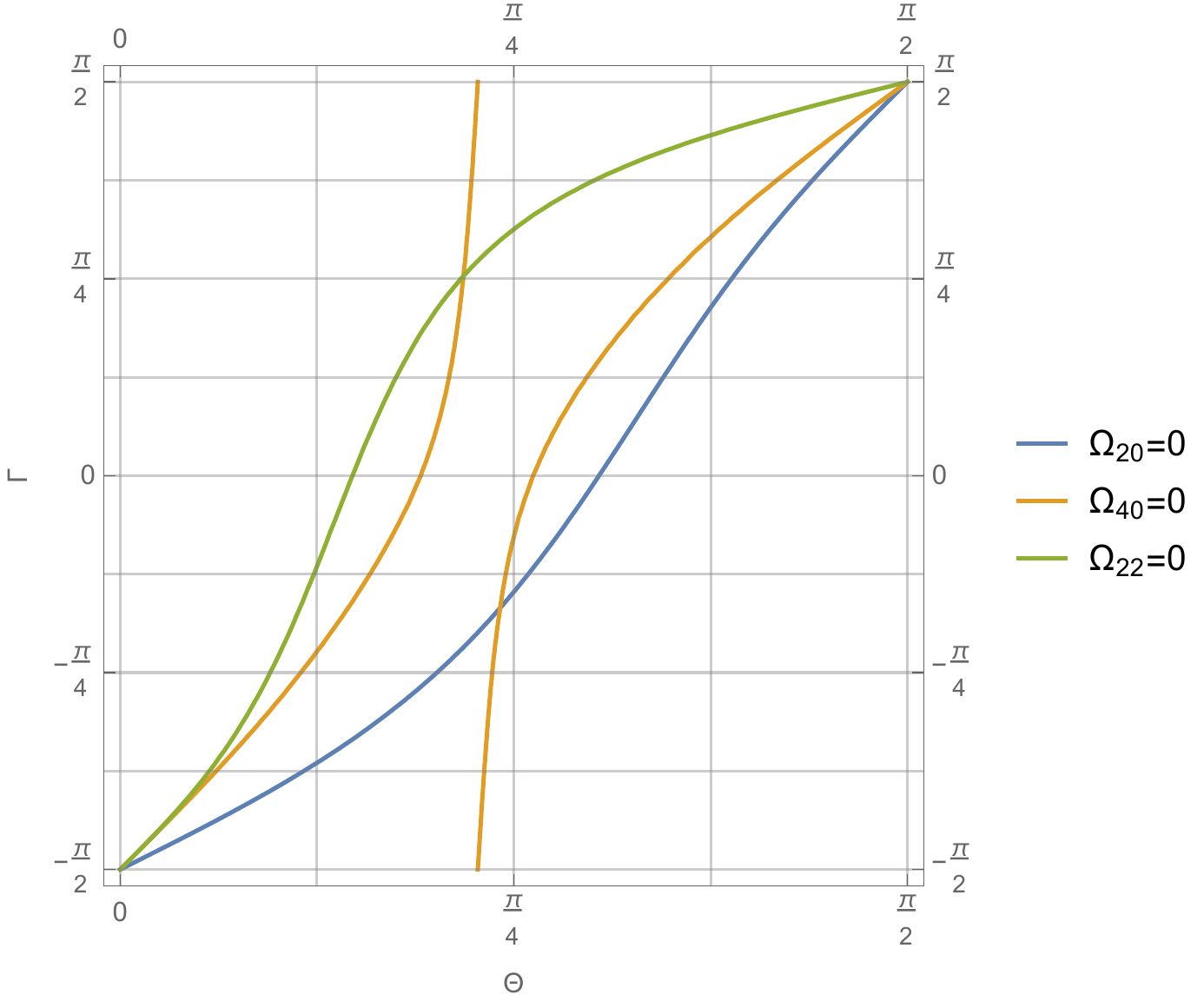}
        \caption{Contours of $\Omega_{20}=0$, $\Omega_{40}=0$, and $\Omega_{22}=0$. There are no non-trivial intersections between $\Omega_{20}$ and $\Omega_{22}$, but there is an intersection between $\Omega_{20}$ and $\Omega_{40}$.}
        \label{fig:4det_cont}
    \end{figure}

\begin{figure}[t]
        \centering
        \includegraphics[width=0.35\textwidth]{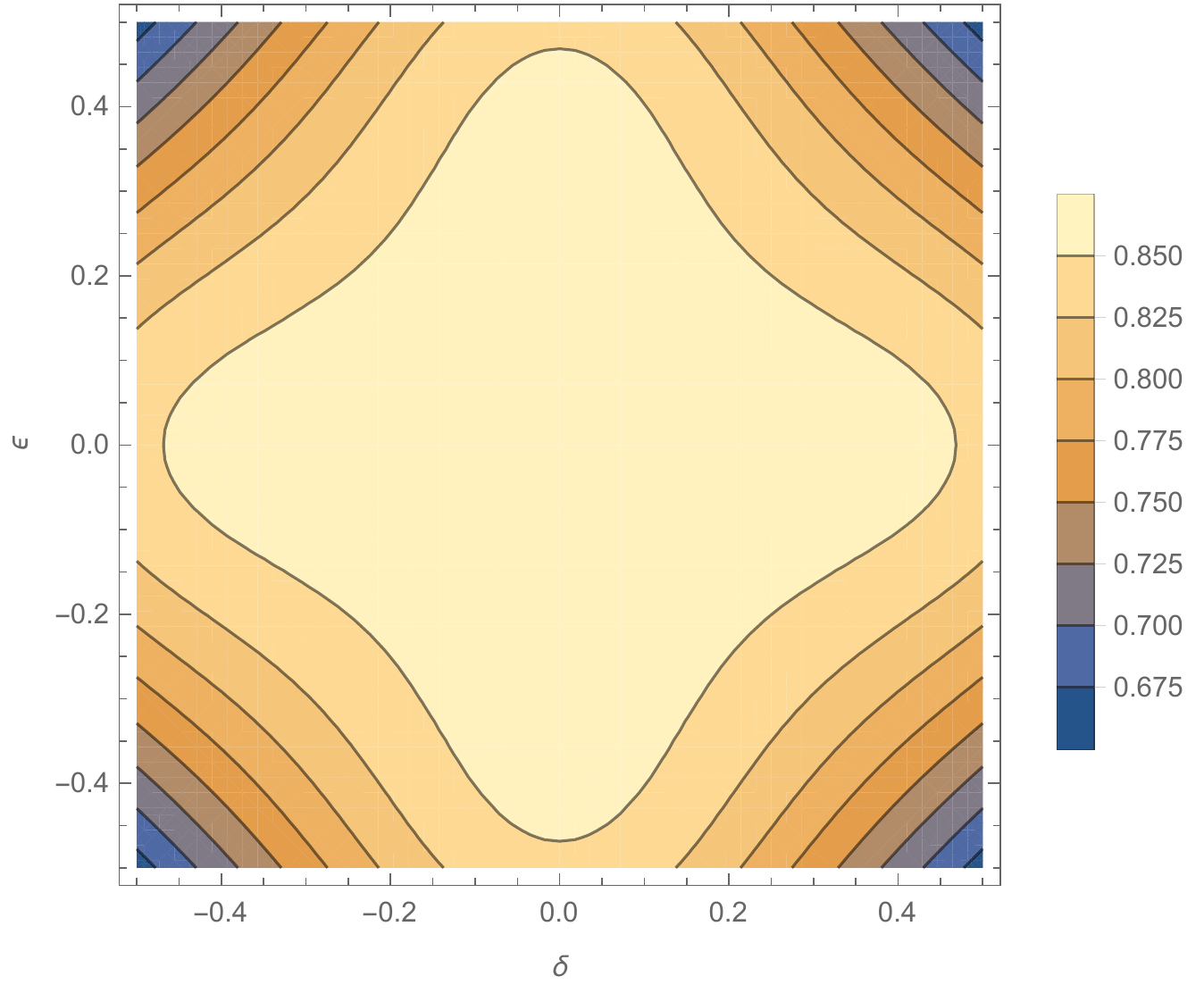}
        \caption{Contour of three dimensional acceptance function, optimized for $\Theta$ and $\Gamma$. Efficiency is more uniform along the direction of the detectors with a change in efficiency of 0.011\% change in efficiency at $\delta$ or $\epsilon=0.2$ along the direction of the detector. There is a change in efficiency of $0.0020\%$ along the detector direction at $\delta$ or $\epsilon=0.15$ and of $0.30\%$ at $\delta=\epsilon=0.15$.}
        \label{fig:4det_opt}
    \end{figure}

    We now consider a detector placed every 90 degrees about the positive $z$-axis. This configuration is three-dimensional (see Fig.~\ref{fig:3D_geometry}) in that the efficiency is a function of the position of the source $(\delta,\epsilon)$ along the plane of the $^6$Li deposit. The average solid angle is given by (see Sec.~\ref{ssec:omega})
    \begin{equation}    \begin{aligned}    \label{eq:3D_omega}
        \Omega(\Theta,\Gamma) = \frac{1}{4} &\left[ \frac{\cos\Gamma - \delta\sin(\Theta+\Gamma)}{(1 + \rho^2-2\delta\sin\Theta)^{3/2}} + \frac{\cos\Gamma + \delta\sin(\Theta+\Gamma)}{(1 + \rho^2+2\delta\sin\Theta)^{3/2}} \right. \\
        &+ \left. \frac{\cos\Gamma - \epsilon\sin(\Theta+\Gamma)}{(1 + \rho^2-2\epsilon\sin\Theta)^{3/2}} + \frac{\cos\Gamma + \epsilon\sin(\Theta+\Gamma)}{(1 + \rho^2+2\epsilon\sin\Theta)^{3/2}} \right]
    \end{aligned}   \end{equation}
    where $\epsilon\equiv y/r$ and $\rho^2\equiv \delta^2+\epsilon^2$.
    Expansion to fourth order in $\delta$ and $\epsilon$, accounting for 4-fold symmetry, introduces one new term, $\Omega_{22}$
    \begin{equation}    \label{eq:4det_Omega_exp_form}
        \Omega= \Omega_0 + \Omega_{20} \cdot (\delta^2+\epsilon^2) + \Omega_{40} \cdot (\delta^4+\epsilon^4) + \Omega_{22} \cdot (\delta^2\epsilon^2).
    \end{equation}
    Like before, the contours of $\Omega_{20}=0$, $\Omega_{40}=0$, and $\Omega_{22}=0$ were plotted, as seen in Fig.~\ref{fig:4det_cont}. There were no non-trivial intersections between $\Omega_{20}$ and $\Omega_{22}$; however, $\Omega_{20}$ and $\Omega_{40}$ intersect at $\Theta=43.44^{\circ}$, $\Gamma= -30.22^{\circ}$.
    Ignoring $\Omega_{22}$ is the only way to simultaneously optimize both $\Theta$ and $\Gamma$. In which case the optimal parameters are $\Theta= 43.44^{\circ}$ and $\Gamma= -30.22^{\circ}$.
    The resulting acceptance function is seen in the contour plot of Fig.~\ref{fig:4det_opt}. Along $\delta$ or $\epsilon$, there is a change in efficiency of $0.0020\%$ at $0.15$ and of 0.30\% at $(\delta=0.15,\epsilon=0.15)$.
    
\section{Non-Uniform $^6$LiF Deposit}   \label{sec:nonuniform}
\begin{figure}[b!]
        \centering
	\includegraphics[width=0.5\columnwidth]{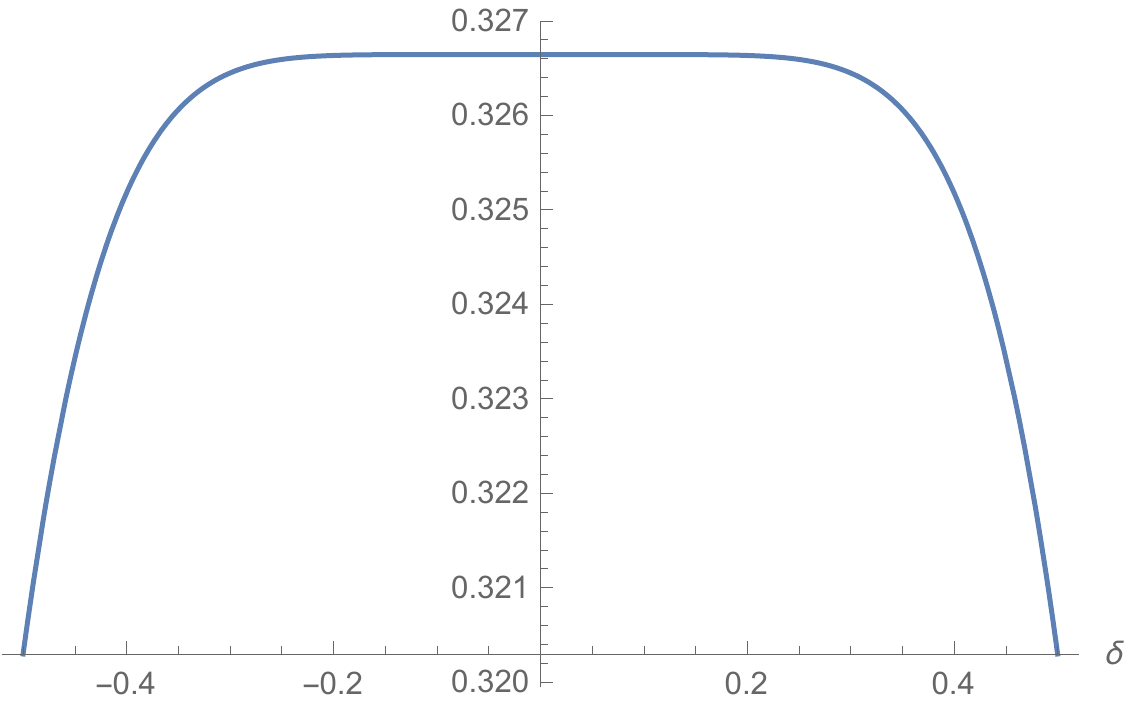}
        \caption{Two detector configuration for $\Theta=55.70^{\circ}$, $\Gamma=70.94^{\circ}$, and $\zeta=0.52$ with a change in efficiency of less than $0.15\%$ until $\delta=0.4$. The change in efficiency at $\delta=0.15$ is around $0.00011\%$.}
        \label{fig:zeta_opt}
    \end{figure}

    Assuming the coating evaporator deposited the $^6$LiF isotropically, there will be a slight non-uniformity of the layer thickness due to the variation in solid angle per area subtended by the deposit as a function of distance from the center. This non-uniformity is parameterized by $\zeta$, where $\zeta r$ is the distance from the substrate to the evaporator during the coating process. The relative thickness $\tau$ of the coating is (see Sec.~\ref{ssec:tau})
    \begin{equation}    \label{eq:tau}
        \tau \equiv \frac{t}{t_{\text{max}}} = \zeta^3 (\zeta^2+\rho^2)^{-3/2}.
    \end{equation}
    The solid angle is scaled by $\tau$ in order to account for the emission rate of particles as a function of position of neutron capture in the foil.
    In the two dimensional case, $\tau\Omega$ is to sixth order
    \begin{equation}
        \tau(\delta;\zeta)\cdot \Omega(\delta;\Theta,\Gamma) = \Omega_{\tau0}(\Theta,\Gamma,\zeta) + \delta^2~ \Omega_{\tau2}(\Theta,\Gamma,\zeta) + \delta^4~ \Omega_{\tau4}(\Theta,\Gamma,\zeta) + \delta^6~ \Omega_{\tau6}(\Theta,\Gamma,\zeta).
    \end{equation}
    $\Omega_{\tau2}$, $\Omega_{\tau4}$, and $\Omega_{\tau6}$ are simultaneously eliminated to give $\Theta=55.70^{\circ}$, $\Gamma=70.94^{\circ}$, and $\zeta=0.52$, with the resulting efficiency seen in Fig.~\ref{fig:zeta_opt}. However, there are values of $\zeta$ that produce comparable uniformity and increased amplitude.
    
\begin{figure}[t!]
        \centering
        \includegraphics[width=0.45\columnwidth]{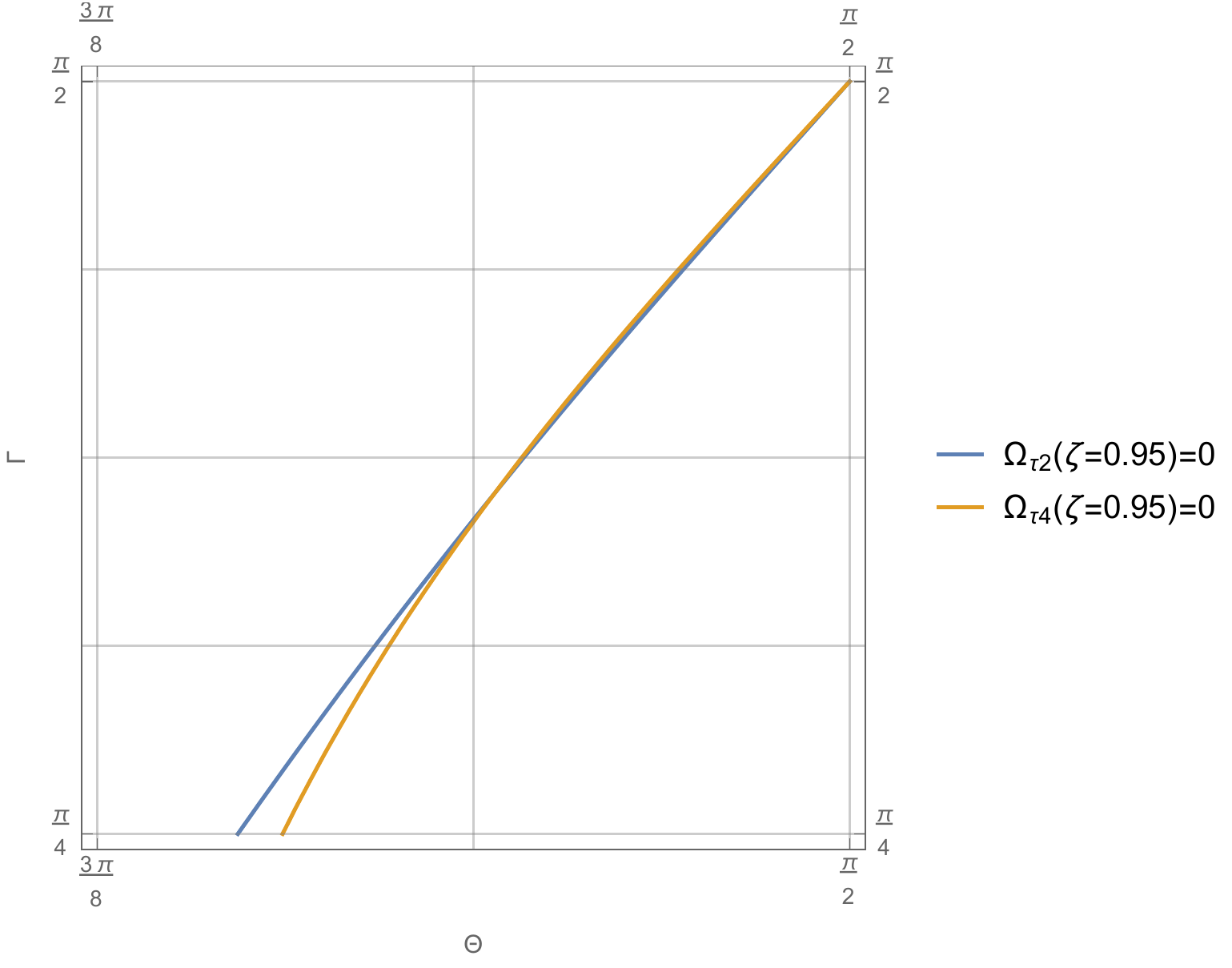}
        \caption{Contours of coefficients at $\zeta=0.95$. An intersection between the $\Omega_{\tau2}=0$ and $\Omega_{\tau4}=0$ contours is now present.}
        \label{fig:zeta_contour}
    \end{figure}

\begin{figure}[b!]
        \begin{minipage}{0.48\textwidth}
            \centering
            \includegraphics[width=0.9\linewidth]{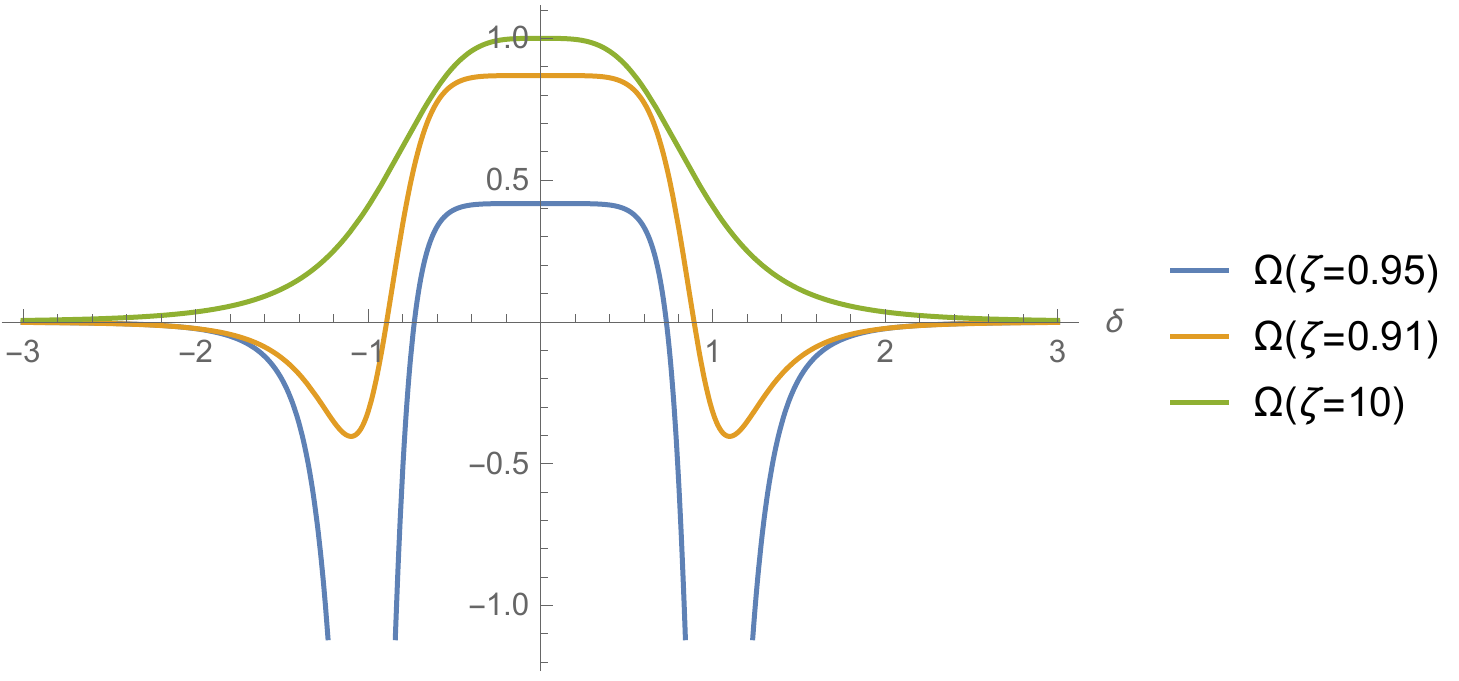}
            \caption{Comparison of large $\zeta$, $\zeta=0.95$, and $\zeta=0.91$ solid angles. Although more uniform about the origin, the solid angle becomes negative for some values of $\delta$.}
            \label{fig:zeta_unstable}
        \end{minipage}\hfill
        \begin{minipage}{0.48\textwidth}
            \centering
            \includegraphics[width=0.9\linewidth]{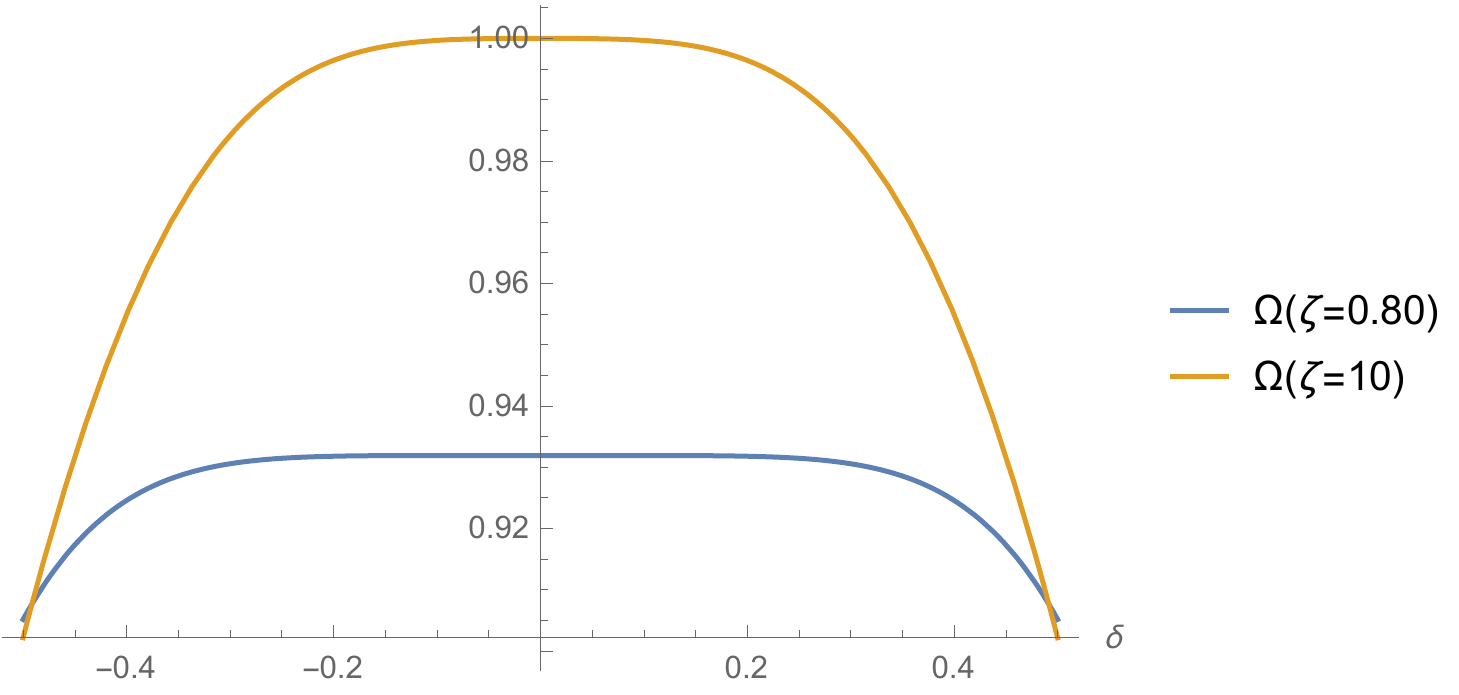}
            \caption{At $\zeta=0.80$, the solid angle had the highest amplitude of all the non-negative results. At $\delta=0.15$, there is a change in efficiency of $0.0022\%$.}
            \label{fig:zeta_stable}
            \end{minipage}
    \end{figure}

    $\Omega_{\tau2}$ and $\Omega_{\tau4}$ can be simultaneously eliminated for $\zeta\leq0.95$ (Fig.~\ref{fig:zeta_contour}). The solid angle becomes negative at large values of $\delta$ indicating the source extended beyond the plane of the near detector. The large optimized tilt angle $\Gamma$ in these cases results in a reduction of total detection efficiency (Fig.~\ref{fig:zeta_unstable}), but around $\zeta=0.8$, $\Omega$ is no longer negative for any $\delta$ and has a change in efficiency of less than 0.73\% up to $\delta=0.4$ and has optimal angles $\Theta=59.37^{\circ}$, $\Gamma=21.27^{\circ}$. This is an improvement from the large $\zeta$ limit which saw a change in efficiency of 0.73\% by $\delta=0.25$ (Fig.~\ref{fig:zeta_stable}).
    
    
    In the three detector configuration, $\Omega_{20}$ and $\Omega_{40}$ remain optimizable at lower $\zeta$'s. However, $\Omega_{20}$ and $\Omega_{22}$ still only give trivial solutions for all $\zeta$. Since an optimal $\Theta$ and $\Gamma$ were already found in the two parameter optimization, this particular analysis does not offer anything advantageous in the four detector configuration.
    
\section{Conclusion} \label{sec:conclusion}
    The results here provide an analysis on minimizing some of the uncertainty associated with neutron flux detector geometries.
    In the two parameter, two-detector optimization, $\Theta=35.26^{\circ}$ and $\Gamma=0^{\circ}$ resulted in sixth order non-uniformities with a maximum change in efficiency of $0.10\%$ over a distance of 3 cm with a point-like detector placed 20 cm away.
    In the three parameter, two-detector optimization, $\Theta=55.70^{\circ}$, $\Gamma=70.94^{\circ}$, and $\zeta= 0.52$ resulted in eighth order non-uniformities with a change in efficiency of $0.00011\%$ over a $\delta$ of 0.15. Although $\zeta=0.8$ ($\Theta=59.37^{\circ}$, $\Gamma=21.27^{\circ}$) provides comparable uniformity ($0.0022\%$) with a higher efficiency amplitude.
    The two parameter three-detector optimization, $\Theta=43.44^{\circ}$ and $\Gamma=-30.22^{\circ}$, resulted in a half-elimination of fourth order non-uniformities with a change in efficiency of $0.0020\%$ over $(\delta=\pm0.15,0)$ or $(0,\epsilon=\pm0.15)$ and $0.3\%$ over $(\pm0.15,\pm0.15)$.
    Full elimination of fourth order non-uniformities may not be possible with the given parameter set, so the three parameter three-detector optimization does not provide any improved uniformity of efficiency.

    The aspect ratio of a rectangular detector (ratio of side-lengths) will provide a new opportunity for optimization. Alternatively, adding a second ring of four detectors with its own set of parameters ($\Theta_2$, $\Gamma_2$) would introduce two more degrees of freedom.
    For BL3, an arrangement of six detectors (two rings of three) may be considered. The unique geometry of such a configuration could provide convenient diagnostic advantages.
    
\section{Appendix}  \label{sec:appendix}
    \subsection{Derivation of Solid Angle}  \label{ssec:omega}
        The solid angle is given by
        \begin{equation}    \label{eq:omega_gen}
            \Omega(\Theta,\Gamma)= \frac{\vec{A}\cdot\vec{d}}{d^3}
        \end{equation}
        where $\vec{d}=\vec{r}-\vec{x}$. For the case of $r\gg a,b$ (dimensions of the detector; see Fig.~\ref{fig:3D_geometry})\footnote{I believe it is useful to have $\vec{A}$ in polar coordinates to make visualizing angles between $\vec{A}$ and $\vec{d}$ simpler.}
        \begin{equation}    \label{eq:area_vec}
            \vec{A}= A (\hat{r} \cos\Gamma - \hat{\Theta} \sin\Gamma).
        \end{equation}
        Along the $\delta$-axis ($\vec{r}=r(\sin\Theta,0,\cos\Theta)$, $\vec{x}=r(\delta,0,0)$), Eq.~\ref{eq:omega_gen} becomes
        \begin{equation}    \label{eq:s2_pre}
            \Omega(\Theta,\Gamma) = \frac{A}{r^2} \frac{(\cos\Gamma\cos\varphi - \sin\Gamma\sin\varphi)}{1 + \delta^2 - 2\delta\sin\Theta}
        \end{equation}
        $\varphi$ is determined using the scalar product, and the solid angle becomes
        \begin{equation}    \label{eq:Omega_2D_result}
            \Omega(\Theta,\Gamma)= \frac{\cos\Gamma - \delta\sin(\Theta+\Gamma)}{(1 + \delta^2-2\delta\sin\Theta)^{3/2}}.
        \end{equation}
        
        In the three-dimensional case, the general solid angle result is given by
        \begin{equation}    \label{eq:3D_omega_gen}
            \Omega= \frac{\cos\Gamma - (\delta\cos\phi+\epsilon\sin\phi) \sin(\Theta+\Gamma)}{(1 + \rho^2-2(\delta\cos\phi+\epsilon\sin\phi) \sin\Theta)^{3/2}}
        \end{equation}
        where $\phi$ is the detector's location about the positive $z$-axis. Both the two and four detector solid angle averages are derived from this expression.
        
    \subsection{Derivation of Relative Thickness Parameter} \label{ssec:tau}
        Imagine a source that sprays $^6$LiF onto the silicon substrate, located a distance $h$ above the substrate surface. Assume the coating evaporator isotropically sends out the $^6$LiF in a region between $\pm x$, resulting in a thickness $t$. This $t$ is defined from a small cylindrical chunk of deposit with volume $dV$ and cross-sectional area $dA$. Using some constants of the system (particle density $\varsigma$, flux $\Phi$) to relate the volume to the solid angle, the thickness is
        \begin{equation}    \label{eq:thickness}
            t = \frac{\Phi}{\varsigma} \frac{d\Omega}{dA}.
        \end{equation}
        As an analogy, picture the ``source-plane'' being a point, and the ``detector'' is the substrate, located directly overhead (with no tilt). By Eq.~\ref{eq:omega_gen}
        \begin{equation}    \label{eq:dOmega/dA}
            \frac{d\Omega}{dA} = \frac{h}{r^3} \frac{1}{(\zeta^2+\rho^2)^{3/2}} = \frac{1}{r^2} \frac{\zeta}{(\zeta^2+\rho^2)^{3/2}}
        \end{equation}
        where $\zeta\equiv h/r$.
        The maximum thickness is when $\rho=0$. The \emph{relative} thickness $\tau$ is then
        \begin{equation}    \label{eq:tau_result}
            \tau \equiv \frac{t}{t_{\text{max}}} = \zeta^3 (\zeta^2+\rho^2)^{-3/2}.
        \end{equation}
        This can be thought of as a ``thickness efficiency'' which will scale the acceptance function.

\printbibliography

\end{document}